# Comparative study of hydrogen embrittlement resistance between additively and conventionally manufactured austenitic stainless steels

Dong-Hyun Lee[*1,2], Binhan Sun[1], Subin Lee[1], Dirk Ponge[1], Eric A. Jägle[1,3], Dierk Raabe[1]

[1]Department for Microstructure Physics and Alloy Design, Max-Planck-Institut für Eisenforschung GmbH, 40237 Düsseldorf, Germany
[2]Department of Materials Science and Engineering, Chungnam National University, 34134 Daejeon, Republic of Korea
[3]Institute of Materials Science, Bundeswehr University Munich, 85579 Neubiberg, Germany

*Corresponding author: dhlee@cnu.ac.kr (D.-H. Lee)

## Abstract

Hydrogen embrittlement in 304L (18wt.% Cr, 8-10wt.% Ni) austenitic stainless steel (ASS) fabricated by laser powder-bed-fusion (LPBF) was investigated by tensile testing after electrochemical hydrogen pre-charging and compared to conventionally available 304L ASSs with two different processing histories, (i) casting plus annealing (CA) and (ii) CA plus thermomechanical treatment (TMT). It was revealed that hydrogen-charging led to a significant reduction in ductility for the CA sample, but only a small effect of hydrogen was observed for the LPBF and CA-TMT samples. Hydrogen-assisted cracking behavior was found to be strongly linked to strain-induced martensitic transformation. In addition, the amount of α′ martensite was much higher in the CA sample than in other samples, suggesting that severe hydrogen embrittlement can be correlated with the low mechanical stability of austenite. Detailed microstructural characterization showed that low austenite stability of the CA sample was mainly attributed to the retained content of δ ferrite and the chemical inhomogeneity inside γ matrix (γ close to δ has ~2 wt.% higher Cr but ~2 wt.% lower Ni), but TMT enhanced the chemical homogeneity and thus austenite stability. By contrast, the LPBF process led directly, i.e. without any thermomechanical treatment, to a fully austenitic structure with homogeneous elemental distribution in the ASS. These results confirmed that the presence of δ and the chemical inhomogeneity inside γ matrix, which promoted the deformation-induced martensitic transformation and the associated H enrichment at the γ-α′ interface, was the primary reason for the severe H-assisted failure.

## 1. Introduction

Austenitic stainless steels (ASSs) are frequently used for sensitive hydrogen (H) storage and hydrogen infrastructure as well transportation applications, since they are generally less susceptible to hydrogen embrittlement (HE) compared to ferritic steels due to their lower diffusivity and higher solubility of H [1–3]. HE describes a phenomenon in which materials undergo an often abrupt and catastrophic deterioration of their mechanical properties (especially when exposed to tensile loads, due to loss of their tensile ductility) caused by the presence of environmental H in acidic solutions and H-containing gases [4–8] which diffuses into the bulk material. In contrast to thermodynamically stable ASSs (e.g., AISI type 310S) which is less prone to undergo HE, severe HE has often been observed in metastable ASSs containing only 8-10 wt.% Ni (e.g., AISI type 304), in which strain-induced $\alpha'$ martensite forms during deformation [9–11]. The strain-induced $\alpha'$ martensite offers a fast diffusion path for H, leading to enrichment of H at critical sites (such as micromechanically highly stressed regions before hetero-interfaces) in the microstructure and thus H-assisted cracking [12,13]. In addition, a small amount of $\delta$ ferrite can be present in metastable ASSs due to segregation during solidification or incomplete $\delta$ to $\gamma$ transformation due to high cooling rates. This may increase HE susceptibility of the samples by providing crack initiation sites [14,15]. There are, however, different reports about the HE susceptibility of metastable ASSs in the literature as summarized in Table 1 [10,13,16–19]. These studies clarify the impact of metallurgical aspects such as grain size, local segregation (caused by different production routes), and volume fraction of $\alpha'$ martensite on the HE susceptibility of 304 ASS. For example, Egels *et al.* [18] showed that 304 ASSs having very similar nominal compositions can exhibit very different HE susceptibilities due to segregation effects resulting from different production routes. In regions where segregation increases austenite stability, the martensitic transformation was suppressed, which caused a slow H uptake/diffusion rate and thus slow crack growth rate. Regarding the impact of pre-deformation on HE susceptibility of metastable ASSs, it has been reported that with increasing pre-deformation levels in 304 ASS, the amount of prior α′ martensite increases and so does the susceptibility for HE [16,20]. However, some conflicting results were also reported [13], arising from the difference in H-charging methods. Detailed information on H charging conditions and the resulting H contents is given in Table 1. Note that some researchers have also found the formation of ε martensite as a precursor for α′ martensite [21–24], however this mechanism has not been reported in the literature on HE of metastable ASSs [10,13,16–19]. We will discuss the possibility of forming ε martensite in the following section.

Recently, additive manufacturing (AM), such as laser powder-bed-fusion (LPBF), has emerged as a promising manufacturing technology, in which three-dimensional parts are printed by progressively adding thin layers of powder material and consolidating them by melting with a laser beam [25–27]. It has become possible for a number of metals (including steels, aluminum alloys, and titanium alloys) to reliably manufacture highly complex or customized parts directly from a computer-coded design file and raw material powder [25–30]. AM processes have thus the potential to replace many conventional production processes.

LPBF-processed ASSs often exhibit highly complex and hierarchical microstructures (e.g., irregular grain structure and solidification substructure) accompanied with elemental segregation due to the combination of high cooling rate (~$10^5$-$10^6$ K/s) and high thermal stress [31–33], which are not accessible through conventional manufacturing routes, such as casting or forging. In this regard, the LPBF process significantly alters the metallurgical aspects described in Table 1 and thus the H effects in LPBF-processed metastable ASSs can be expected to be very different from those in their conventional counterparts. Accordingly, the analysis of the H effects in LPBF-processed metastable ASSs is essential for examining the possibility of the future use of LPBF-manufactured parts in infrastructures with strong H exposure. Until now, only little systematic research on the issue is available in the literature [34]. The authors in Ref. [34] reported that AM-processed 304L was less susceptible to HE than wrought 304L, and speculated that high HE resistance in AM-processed 304L was attributed to its high austenite stability. However, the questions of the exact role of AM-induced microstructural change in austenite stability and HE resistance still remain unanswered.

As the first step to shed light on this issue, here we systematically investigate HE in LPBF-processed 304L ASS through tensile testing with and without electrochemical H pre-charging. To elucidate the effect of LPBF-induced microstructural change on HE susceptibility, conventionally available 304L ASSs subjected to two different processing histories was also investigated: one is casting plus annealing (CA) and the other is CA plus thermomechanical treatment (TMT).

Table 1. Comparison of metallurgical aspects, H-charging conditions, and resultant HE index in various 304 ASSs from literature. HE index represents the ratio of the elongation (or reduction of area) of the H charged sample, $\varepsilon_{f,H}$, to that of the uncharged sample, $\varepsilon_{f,0}$, which are measured from tensile tests at room temperature. Note that HE susceptibility increases with decreasing HE index ($\varepsilon_{f,H}/\varepsilon_{f,0}$). H charging was conducted either before tensile test or during the test, which are represented as Pre- and In-situ, respectively. H content was measured by thermal desorption spectrometry; the values marked with * mean H content exclusively at brittle H affected zone. Detailed information can be found in each reference.

| Metallurgical aspect | H charging condition | | | | | H content [wt.ppm] | HE index | Ref. |
|---|---|---|---|---|---|---|---|---|
| Heat treatment | Charging method | $H_2$ pressure [MPa] | Current density [mA/cm$^2$] | Time [h] | Temp. [°C] | | | |
| Solution annealed | Gaseous H charging | 1 | - | Duration of tensile test (In-situ) | RT | - | ~0.53 | [9] |
| Sensitized | | | | | | | ~0.29 | |
| Sensitized + desensitized | | | | | | | ~0.51 | |
| Production route | | | | | | | | |
| Casting | Gaseous H charging | 40 | - | Duration of tensile test (In-situ) | RT | - | ~0.50 | [18] |
| Electroslag remelting | | | | | | | ~0.89 | |
| Grain size [μm] | | | | | | | | |
| 12 | Electro-chemical H charging | - | 10 | Duration of tensile test (In-situ) | RT | - | ~0.80 | [19] |
| 8 | | | | | | | ~0.84 | |
| 4 | | | | | | | ~0.95 | |
| Single crystal | Electro-chemical H charging | - | 2.7 | 7 (Pre-) | 80 | - | ~0.43 | [17] |
| 0.46 | | | | | | 124 | ~0.67 | |
| 0.32 | | | | | | 126 | ~0.86 | |
| 0.14 | | | | | | 170 | ~0.84 | |
| 0.09 | | | | | | - | ~0.80 | |
| Pre-strain % (prior α′ vol.%) | | | | | | | | |
| 0 (0) | Gaseous H charging | 30 | - | 48 (Pre-) | 200 | ~48 | ~0.67 | [13] |
| 12 (31) | | | | | | ~39 | ~0.79 | |
| 20 (60) | | | | | | ~22 | ~0.92 | |
| 30 (96) | | | | | | ~6 | ~0.99 | |
| 0 (0) | Electro-chemical H charging | - | 50 | 96 (Pre-) | RT | ~606.5* | ~0.93 | [16] |
| 3 (1) | | | | | | - | ~0.93 | |
| 6 (3) | | | | | | - | ~0.91 | |
| 10 (8) | | | | | | ~834.8* | ~0.82 | |
| 15 (20) | | | | | | - | ~0.71 | |
| 20 (37) | | | | | | ~580.7* | ~0.65 | |
| 25 (57) | | | | | | - | ~0.30 | |

## 2. Materials and experiments

Gas-atomized spherical 304L ASS powder (~45 μm D50, provided by LPW Technologies) with the nominal composition listed in Table 2 was used in this study. A block of 304L ASS with the size of 20 mm (width) × 60 mm (length) × 30 mm (thickness) was fabricated by LPBF system (Aconity Mini, Aconity 3D GmbH, Aachen, Germany). During the LPBF process, the layer thickness, laser power and scan speed were set at 30 μm, 180 W, and 700 mm/s, respectively. The laser scanned line by line along the same direction in each layer with hatch spacing of 80 μm, and the scanning direction was rotated by 90° after each layer. The sample was built up by repeating this process and referred to as the LPBF sample.

The conventional 304L ASS was produced by casting and subsequent solution annealing at 1050 °C for 30 min (hereafter, designated the CA sample). To clarify the effect of TMT in the ASS, a part of the material was, after casting and annealing, subjected to hot-rolling at 1000 °C to a thickness reduction of 50% and subsequently homogenized at 1150 °C for 18 hours in Ar atmosphere. Then, the sample was finally cold-rolled to a thickness reduction of ~70% and annealed at 1050 °C for 1 hour, followed by water quenching. The CA sample subjected to TMT will be referred to as the CA-TMT sample. Thus, there are three different samples that were tested here; samples LPBF, CA, and CA-TMT. The compositions of samples LPBF, CA and CA-TMT were analyzed by inductively coupled plasma optical emission spectrometry (ICP-OES) and the results are also listed in Table 2. Due to the different production routes, the LPBF and the CA(-TMT) samples vary slightly in chemical composition.

To estimate the different austenite stability of all samples due to these slight compositional differences, the thermodynamic driving force for α′ martensite transformation was estimated by calculating $\Delta G^{\gamma\rightarrow\alpha}$ (=$G_\alpha$–$G_\gamma$, where $G_\gamma$ and $G_\alpha$ are Gibbs free energies of γ austenite and α ferrite, respectively) using the Thermo-Calc software, version 2018a, in conjunction with the TCFE9 database. The calculations were carried out for a temperature of 298 K considering the chemical compositions shown in Table 2. The values of $\Delta G^{\gamma\rightarrow\alpha}$ for samples LPBF and CA(-TMT) are calculated as -2.66 kJ/mol and -2.57 kJ/mol, respectively, i.e. the austenite in the CA(-TMT) sample is only very slightly more stable than that in the LPBF sample. In addition, the driving force for ε martensite transformation, $\Delta G^{\gamma\rightarrow\varepsilon}$ (=$G_\varepsilon$–$G_\gamma$, where $G_\varepsilon$ is Gibbs free energy of ε), was also calculated. $\Delta G^{\gamma\rightarrow\varepsilon}$ values (-0.50 kJ/mol and -0.61 kJ/mol for samples LPBF and CA(-TMT), respectively) are much lower than $\Delta G^{\gamma\rightarrow\alpha}$ values, indicating that α′ martensite transformation is thermodynamically more favorable in the present 304L ASSs. As a second approach to estimate the austenite stability, the $M_{d30}$ temperature, which represents the temperature at which 50 vol.% of the austenite transforms to martensite with an applied true strain of 30%, is calculated by using the following empirical expression [35]:

$$M_{d30}(°C) = 413 - 462(C + N) - 9.2Si - 8.1Mn - 13.7Cr - 9.5Ni - 18.5Mo, \quad (1)$$

where all elements are counted in weight percentage (wt. %). The $M_{d30}$ temperature serves as an approximate parameter representing the mechanical stability of the austenite against deformation-inducted martensite formation; a lower $M_{d30}$ temperature means a higher mechanical stability of austenite. Based on the chemical composition in Table 2 and Eq. (1), $M_{d30}$ of samples LPBF and CA(-TMT) were calculated to be ~19.1 °C and ~13.6 °C, respectively. Again, this indicates only a very slightly higher austenite stability of the CA(-TMT) specimen compared to the LPBF specimen. Considering these insignificant differences in $\Delta G^{\gamma\rightarrow\alpha}$ and $M_{d30}$ between samples LPBF and CA(-TMT), the austenite stability of LPBF can be considered as very similar to that of CA(-TMT), which suggests that the CA(-TMT) sample can serve as a reference material in this study.

Table 2. Chemical compositions (wt.%) of 304L ASSs investigated in this study.

| Material | C | Cr | Cu | Mn | N | Ni | O | P | S | Si | Fe |
|---|---|---|---|---|---|---|---|---|---|---|---|
| Powder | 0.017 | 18.4 | <0.1 | 1.3 | 0.060 | 9.90 | 0.03 | 0.018 | 0.004 | 0.63 | Bal. |
| LPBF | 0.017 | 18.3 | 0.029 | 1.2 | 0.055 | 9.94 | 0.03 | 0.002 | 0.007 | 0.63 | Bal. |
| CA(-TMT) | 0.025 | 18.2 | 0.453 | 1.87 | 0.075 | 9.01 | - | 0.031 | 0.018 | 0.34 | Bal. |

Flat, dog-bone-shaped tensile samples with gauge dimensions of 2 mm (width) × 4 mm (length) × 1mm (thickness) were machined by spark erosion from the LPBF-produced block and mechanically ground. Tensile specimens were cut from the horizontal section (XY plane) perpendicular to the building direction (BD) as shown in Fig. 1. In order to introduce

H into the specimen, electrochemical H-charging was carried out at room temperature with a potentiostat using a 2 g/l $CH_4N_2S$ (hydrogen recombination agent) in 0.5 M $H_2SO_4$ solution (pH = 1). H-charging was performed under two different conditions; one set of samples was H pre-charged for 1 day at a constant current density of 20 $mA/cm^2$, and another set was H pre-charged at 50 $mA/cm^2$ for 5 days. To minimize the influence of outgassing, all further experiments on H pre-charged samples were started within 30 min after H-charging. Tensile tests were performed on both uncharged and H pre-charged samples using a tensile stage (Kammrath & Weiss, Dortmund, Germany) at a cross-head speed of 0.4 μm/s, which corresponds to an initial strain rate of $10^{-4}$ /s. The digital image correlation (DIC) technique was carried out to determine the local plastic strain during tensile deformation and the obtained data was analyzed using Aramis software (GOM GmbH, Braunschweig, Germany). At least three tests were conducted for each condition to confirm reproducibility.

The microstructure of the specimens was characterized by an EBSD detector (Digiview, EDAX-TSL, Mahwah, NJ) mounted on a scanning electron microscope (SEM; JSM 6500F, JEOL Ltd., Tokyo, Japan), electron channeling contrast imaging (ECCI; Merlin, Carl Zeiss AG, Oberkochen, Germany), and transmission electron microscopy (TEM; JEM 2200FS, JEOL Ltd., Tokyo, Japan) in scanning TEM (STEM) mode. For SEM probing, the samples were mechanically ground with fine SiC papers (grit number up to 2000) and subsequently polished with 3 μm diamond suspension and finally with 0.05 μm silica suspension. TEM specimen was prepared by focused ion beam (FIB) milling with a Nova 600 NanoLab (FEI Co., Hillsboro, OR). During SEM and TEM analysis, element mapping was performed using energy dispersive X-ray spectroscopy (EDS).

The H desorption behavior was analyzed by using a custom-designed UHV-based thermal desorption spectroscope (TDS) equipped with a mass spectrometer detector. During the TDS measurements, the samples were heated up to 800 $^oC$ at a constant heating rate of 26

K/min. The total H content was determined by measuring the cumulative amount of the desorbed H from room temperature to 800 °C. Note that the specimens for TDS were also cut from the horizontal plane (see Fig. 1).

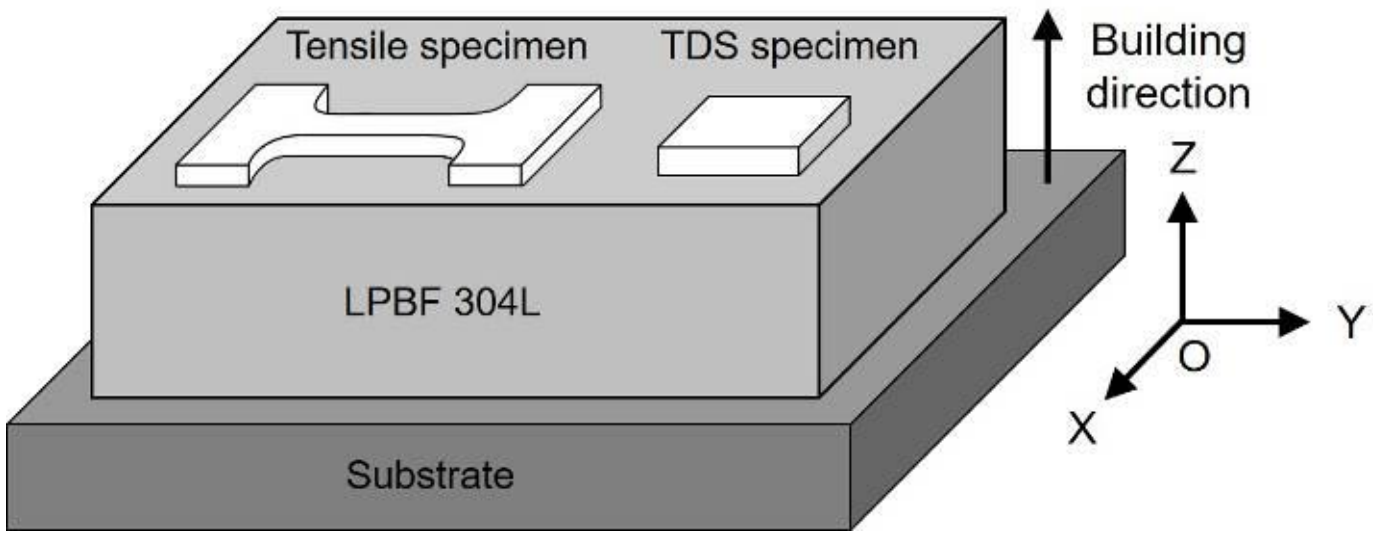

Fig. 1 Schematic illustration of the samples for tensile test and TDS measurement. (LPBF: Laser powder-bed-fusion, TDS: Thermal desorption spectroscope).

## 3. Results

### 3.1 Initial microstructure and the effects of H on the tensile properties

The microstructural characterization of the LPBF sample by EBSD and ECCI is shown in Fig. 2. The inverse pole figure (IPF) map of the LPBF sample (Fig. 2a) reveals that the grains of the sample exhibit a highly irregular shape with a wide range of sizes (~1–80 μm) which is attributed to the track-wise melting of material and the high temperature gradient during the solidification. The average grain size (counting only high-angle grain boundaries (HAGBs)), is estimated to be ~47 μm measured by using the linear intercept method. The LPBF sample consists of a single-phase austenite structure with a large fraction of low-angle grain boundaries (LAGBs, ~60% of the total GBs, Fig. 2b). Furthermore, Fig. 2c displays an ECCI image of the LPBF sample, showing that a solidification cellular structure is embedded in individual grains. The size of this cellular structure is ~350 nm in diameter (which is substantially smaller than that of the grains) and a high density of dislocations are concentrated at the wall of the cells (Fig. 2d), similar to the cell structures reported in Al-, Co-, and Fe-based samples fabricated by LPBF process [31,32,36]. When solidification cells grow together into coarse single grains under a high temperature gradient and at high growth rate conditions, the slight orientation difference between neighboring cells causes the high density of such dislocation walls [32,37].

In contrast to the LPBF sample, the IPF map of the CA sample (Fig. 3a) shows regular polygonal grains with a large fraction of annealing twins. The estimated average grain size of the CA sample is ~51 μm (including twin boundaries). More importantly, retained δ ferrite is present in the γ austenite matrix (Fig. 3b), which is not observed in the LPBF sample, and its volume fraction is ~1vol.%. This implies that the solution annealing at 1050 °C for 30 min is insufficient to remove δ formed during casting. Fig. 3c and 3d display the IPF map and the corresponding phase map for the same area of the CA-TMT sample, respectively. It is evident that the CA-TMT samples has a slightly larger grain size (~62 μm) than the CA sample (Fig. 3c) and, more importantly, TMT completely eliminates the retained δ ferrite (Fig. 3d). These results clearly suggest that the 304L sample produced by standard casting requires an additional TMT for achieving fully austenitic structure, in contrast to the 304L sample produced by LPBF.

In summary, the results presented in Fig. 2 and Fig. 3 demonstrate that large temperature gradients and rapid solidification as imposed during the LPBF process imparts three crucial changes to the microstructure in 304L ASS compared to conventional materials (especially compared to the CA sample): (i) fully austenitic structure without any further processing step, (ii) irregular shaped grains with a wide range of sizes, and (iii) the presence of a large fraction of LAGBs and the associated cellular substructure within the individual grains.

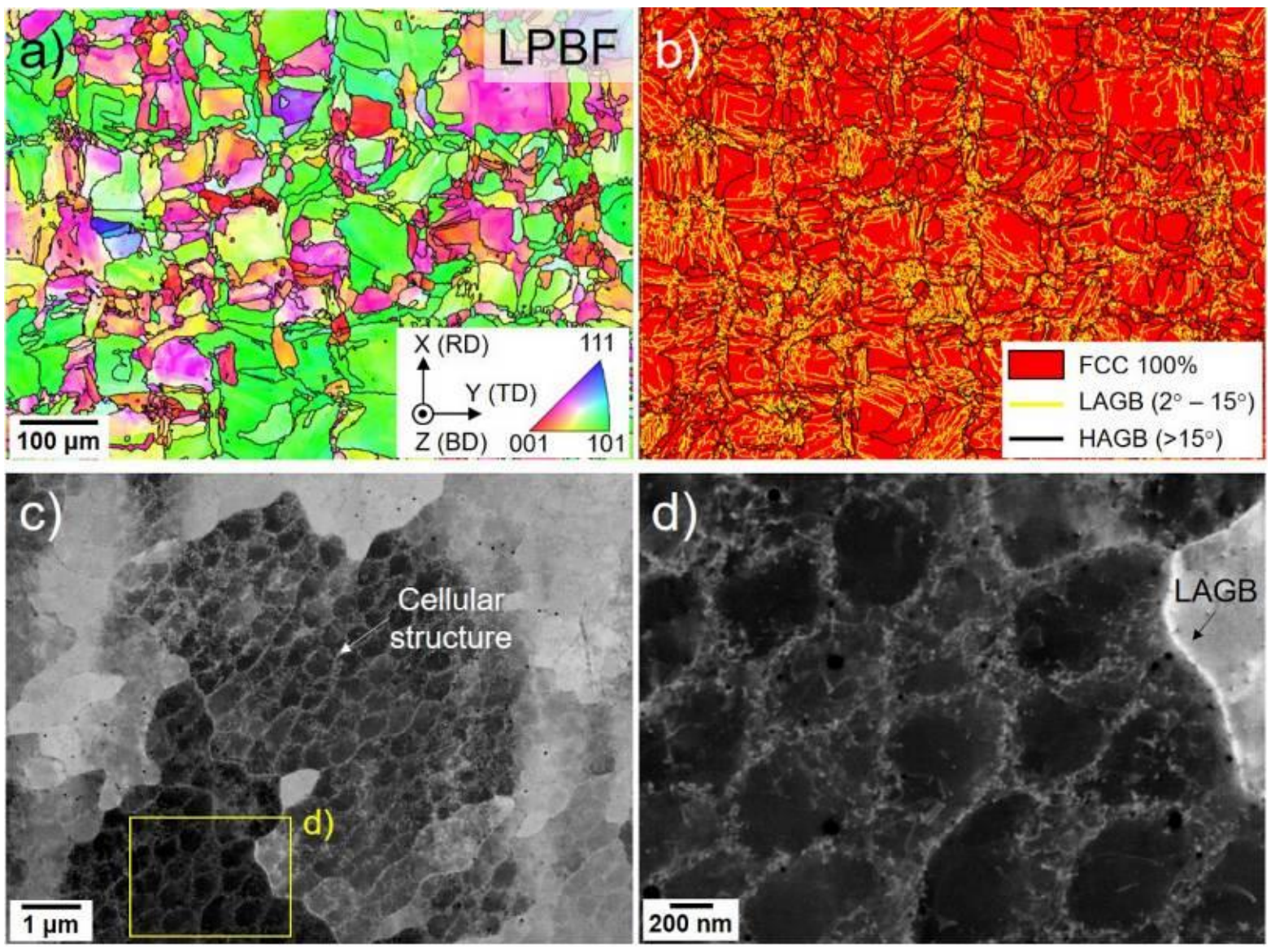

Fig. 2 Microstructure of the LPBF sample: (a) EBSD inverse pole figure map, (b) EBSD phase map with LAGBs and HAGBs superimposed, (c) ECCI image showing cellular structure, and (d) magnified ECCI image of selected area in (c). (LPBF: Laser powder-bed-fusion).

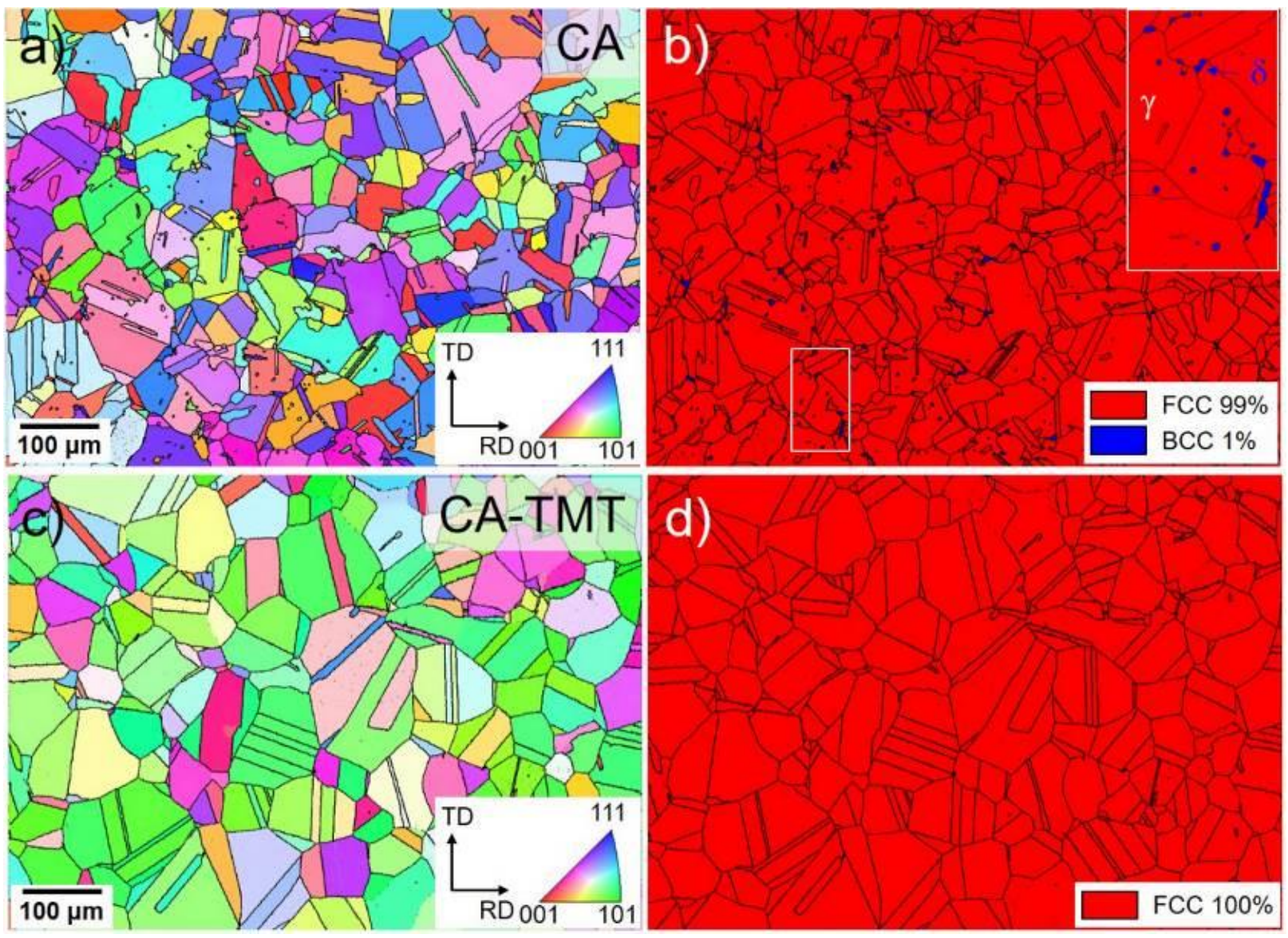

Fig. 3 Microstructure of samples (a,b) CA and (c,d) CA-TMT: (a) and (c) EBSD inverse pole figure map, (b) and (d) EBSD phase map. (CA: Casting plus annealing, TMT:

Thermomechanical treatment).

The representative engineering stress vs. engineering strain responses of samples LPBF, CA, and CA-TMT, as well as their H-charged counterparts (charged for 1 and 5 days) are shown in Fig. 4. Before H-charging, the LPBF sample exhibits higher yield strength (~520 MPa) compared to that of the CA and CA-TMT samples (~340 MPa and ~240 MPa, respectively) due to the high density of LAGBs and the solidification cell structure which acts as an effective obstacle to dislocation motion [31,32]. However, the LPBF sample shows lower elongation to failure (~62%) than the other materials (~85% and ~96% for the CA and CA-TMT samples, respectively), which is attributed to the combined effects of the general trade-off between strength and ductility, the high dislocation density, and the presence of LPBF-induced defects like voids and pores [27,38,39]. After H-charging for 1 day, the ductility of all 304L AASs gets reduced, as expected (Fig. 4). However, each sample exhibits different HE susceptibility: H-charging results in a significant loss in ductility for the case of the CA sample, whereas only a small effect of H on ductility is observed for the LPBF and CA-TMT samples. With further increasing H-charging time to 5 days, a larger loss in ductility was observed for all investigated samples, but the LPBF and CA-TMT samples are still less susceptible to HE than the CA sample. The HE susceptibility of a material is usually quantitatively represented as the ratio of the total elongation to failure of the H-charged sample ($\varepsilon_{f,H}$) to that of the uncharged sample ($\varepsilon_{f,0}$), i.e., $\varepsilon_{f,H}/\varepsilon_{f,0}$ [9,13]. As shown in Fig. 4d, the values of $\varepsilon_{f,H}/\varepsilon_{f,0}$ obtained from the CA-TMT and LPBF samples are obviously higher than those obtained from the CA sample. It is also noted that the H-charged CA sample shows a large number of cracks on the surface after tensile testing, but the CA-TMT and LPBF samples show relatively less H-assisted cracking (see insets in Fig. 4). These results suggest that HE susceptibility of the present 304L ASSs is apparently related to the presence of retained δ.

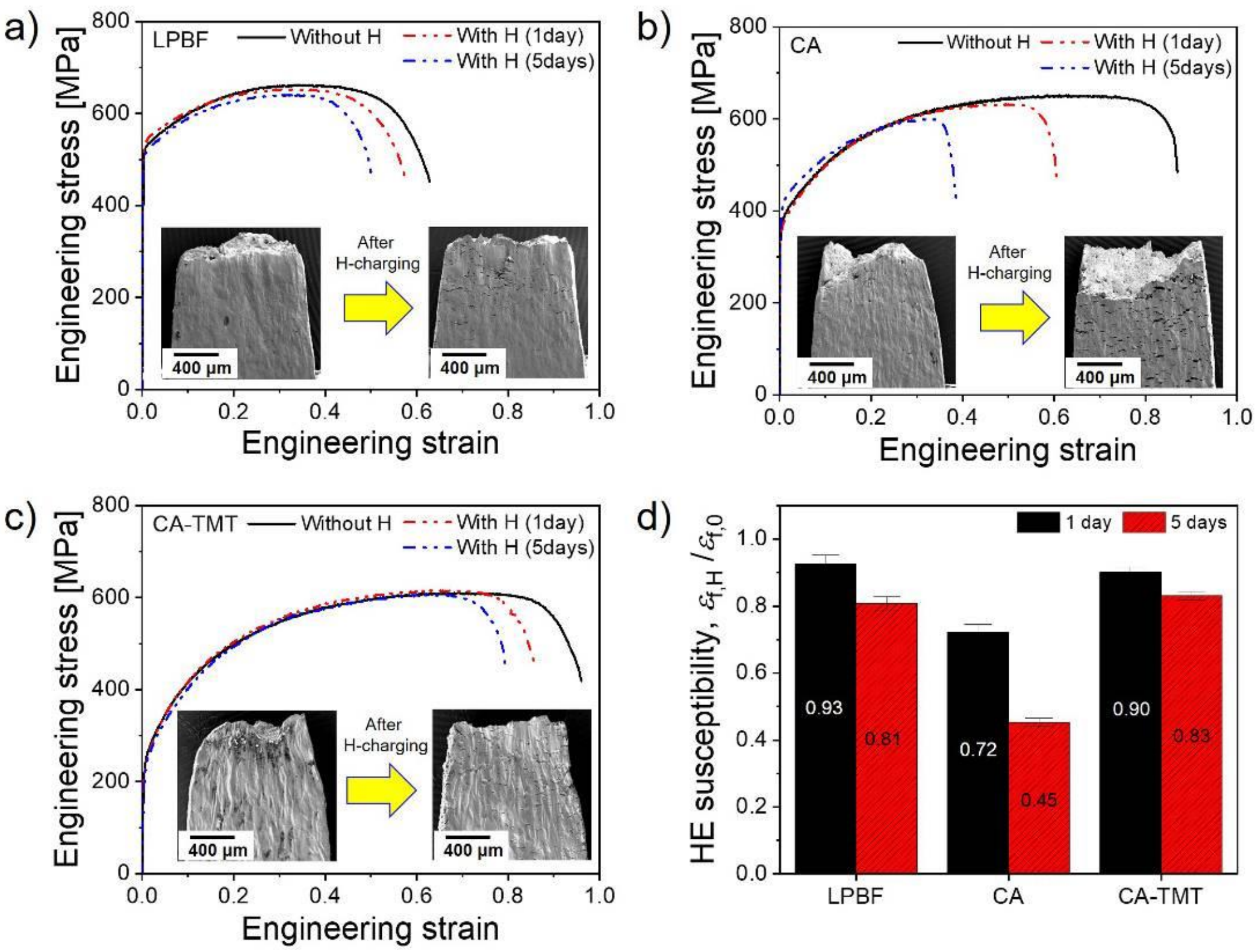

Fig. 4 Typical engineering stress-engineering strain curves of the (a) LPBF, (b) CA, and (c) CA-TMT samples with and without H. The insets show the SEM images of the surface of samples after tensile testing at the region of fracture. (d) Quantification of HE susceptibility for each condition. Error bars represent the standard deviations of the $\varepsilon_{f,H}/\varepsilon_{f,0}$ values obtained from several stress-strain curves. (LPBF: Laser powder-bed-fusion, CA: Casting plus annealing, TMT: Thermomechanical treatment).

The H contents in the samples after H-charging were measured by analyzing TDS curves, and the results are shown in Fig. 5. The inset of Fig. 5 displays the variation in H content with H-charging time for the LPBF sample, which suggests that H content gradually increases with H-charging time. After 5 days H-charging (main plot of Fig. 5), all samples show a single desorption peak at similar temperatures (~220 $^{o}$C), indicating that the H trapping sites are similar in the present samples. In general, such a low-temperature peak corresponds to relatively weak H trapping sites such as interstitial lattice sites and weak H trapping defects such as dislocations and GBs [40–42]. It is also interesting to note that, although the LPBF sample contains a large fraction of LAGBs and cellular walls, the total H content in the LPBF sample is very similar to that in the CA-TMT sample (see Fig. 5). In addition, despite showing a significant HE effect, the CA sample exhibits a slightly lower H content than the other samples. On this basis, we can conclude that the presence of LAGB and cellular walls has a negligible effect on the H content of the LPBF sample and, more importantly, the different HE susceptibilities of the present samples are not primarily related to their total H contents.

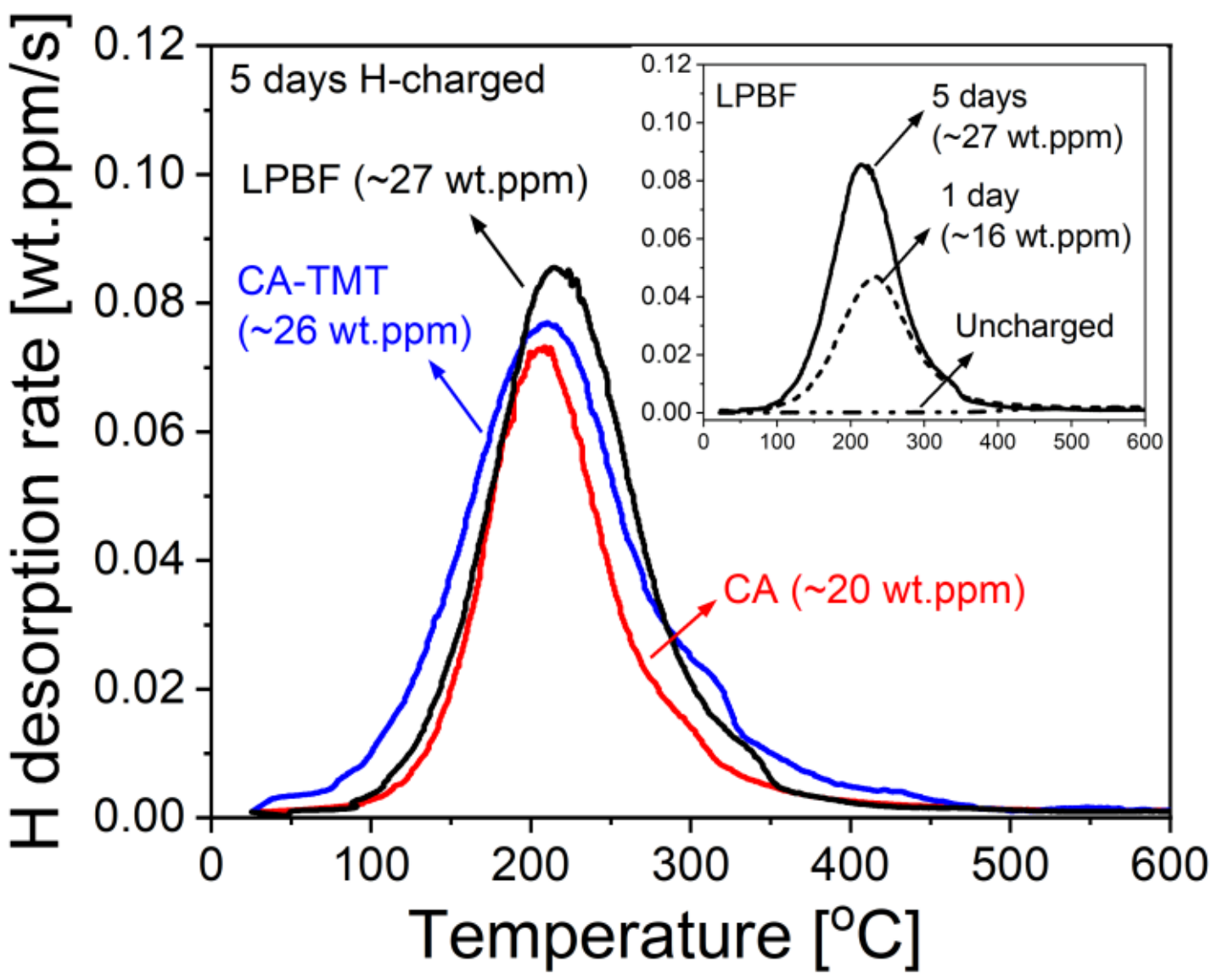

Fig. 5 H desorption curves of the 5 days H-charged samples at a heating rate of 26 °C /min. The inset shows the variation in H desorption curve with H-charging time for the LPBF sample. (LPBF: Laser powder-bed-fusion, CA: Casting plus annealing, TMT: Thermomechanical treatment).

3.2 Deformation microstructure without the presence of H

With the aid of DIC mapping, the evolution of the deformation microstructure in the uncharged samples was investigated by ECCI on different regions corresponding to different local strain levels (from ~5% up to ~40%). At a low strain of ~5%, deformation twins with two different twinning systems are observed (Fig. 6a) which are nucleated from HAGBs and penetrate through the cellular walls. From the zoom-in image of 5% strained microstructure (middle of Fig. 6a), a large density of stacking faults (SFs) can be observed within a cell, indicating that dislocation glide is mainly governed by partial dislocations, whose motion is impeded by the cellular walls [32]. With increasing strain to ~40%, a significant increase in the mechanical twin density is observed for both twinning systems.

In contrast, the CA sample shows a different microstructural evolution during deformation (Fig. 6b). A low strain of ~5% resulted in substantial increase in the density of dislocations. At this stage, multiple intersecting planar slip lines are observed, suggesting that the plastic deformation mainly occurs by planar slip. We identified that the slip lines are along {111} planes by EBSD trace analysis. Upon increasing the strain level to ~20%, a well-developed dislocation cell structure is observed and some grains show mechanical twins with a thickness of a few nanometers. A further increase in the strain to ~40% leads to activation of two different twinning systems and a significant increase in mechanical twin density for both twinning systems. This observation is in agreement with prior reports on deformation mechanisms of conventional ASSs, i.e. the co-existence of dislocation slip and mechanical twinning [43–45]. The development of deformation microstructure in the CA-TMT sample was found to be similar to that in the CA sample, except for the absence of δ. Thus, only the results for the CA samples are displayed in Fig. 6.

These microstructure observations demonstrate that dislocation glide in the LPBF sample is more difficult compared to samples CA and CA-TMT due to the presence of the cellular structure and thus mechanical twinning was favored and activated already at an early stage of deformation. However, as shown in Fig. 6, there is no significant difference in the deformed microstructure between samples LPBF and CA(-TMT) at relatively high strain levels (above ~20 %). This observation suggests that the intrinsic material properties such as the stacking fault energy may be similar for both samples, but that the critical stress for twinning is reached at an earlier stage of deformation in the LPBF sample because of its high yield stress resulting from the presence of the cellular structure. Another interesting observation is that the cellular structure in the LPBF sample remained stable during the entire deformation (Fig. 6a).

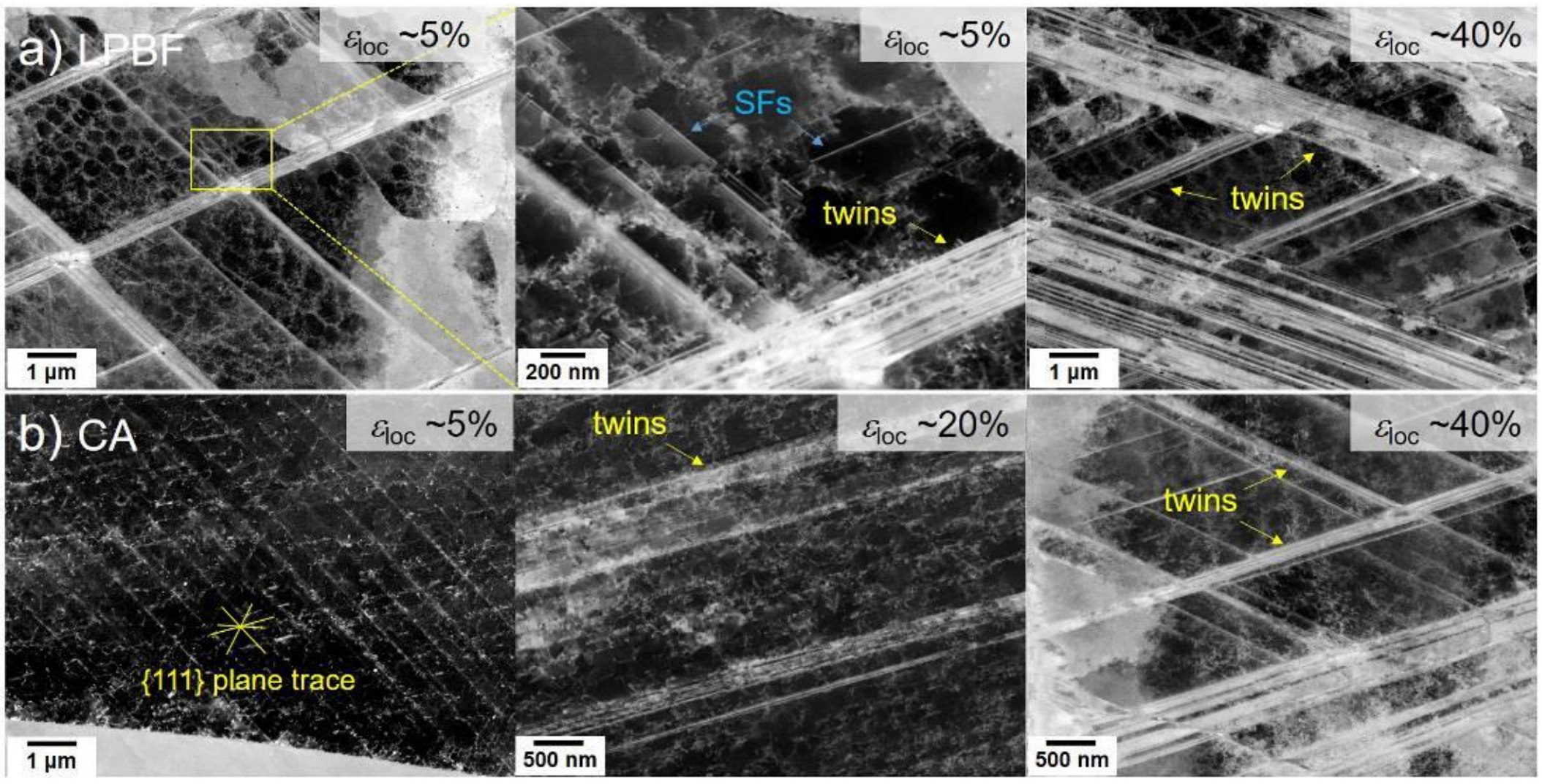

Fig. 6 Deformation microstructure of the (a) LPBF and (b) CA samples revealed by ECCI at different strain levels. (LPBF: Laser powder-bed-fusion, CA: Casting plus annealing).

According to the values of $\Delta G^{\gamma\rightarrow\alpha}$, all investigated samples have a metastable austenitic matrix, which conforms to the presumption of a minimum driving force of -2.1 kJ/mole that is necessary for deformation-induced martensitic transformation [46]. However, it is hard to distinguish strain-induced α′ martensite from the γ matrix in ECCI images (see Fig. 6). Therefore, the microstructure evolution of the uncharged samples with increase in local strain from 20% to 60% was investigated also by EBSD mapping. Fig. 7a-c display EBSD phase maps for the LPBF, CA, and CA-TMT samples, respectively. All samples exhibit similar nucleation behavior and distribution of α′ martensite; rod-shaped α′ martensite is mostly nucleated at the intersections of mechanical twins, and developed within the individual twins, which is in good agreement with previous studies on strain-induced martensitic transformation in ASSs (i.e., γ→twinning→α′) [21–24]. Although γ→ε→α′ is another possible transformation sequence [21–24], α′ martensite transformation in the present 304L ASSs most likely occurs in the sequence γ→twinning→α′ due to the substantially low $\Delta G^{\gamma\rightarrow\varepsilon}$ values (-0.50 kJ/mol). Interestingly, the volume fraction of α′ ($f_{\alpha'}$) in the CA sample is much higher than that in the LPBF and CA-TMT samples at the same local strain. For example, $f_{\alpha'}$ at a local strain of 60% for samples LPBF, CA, and CA-TMT are ~3%, ~16%, and ~1%, respectively (Fig. 7). Note that the CA sample shows both δ and α′, but δ can be distinguished from α′ in terms of its blocky morphology (see Fig. 7b). It is somewhat interesting to note that, although all samples show similar $\Delta G^{\gamma\rightarrow\alpha}$ and $M_{d30}$ values, the CA sample shows a much lower austenite stability than the others. The possible explanation for this unexpected behavior will be provided in section 4.1.

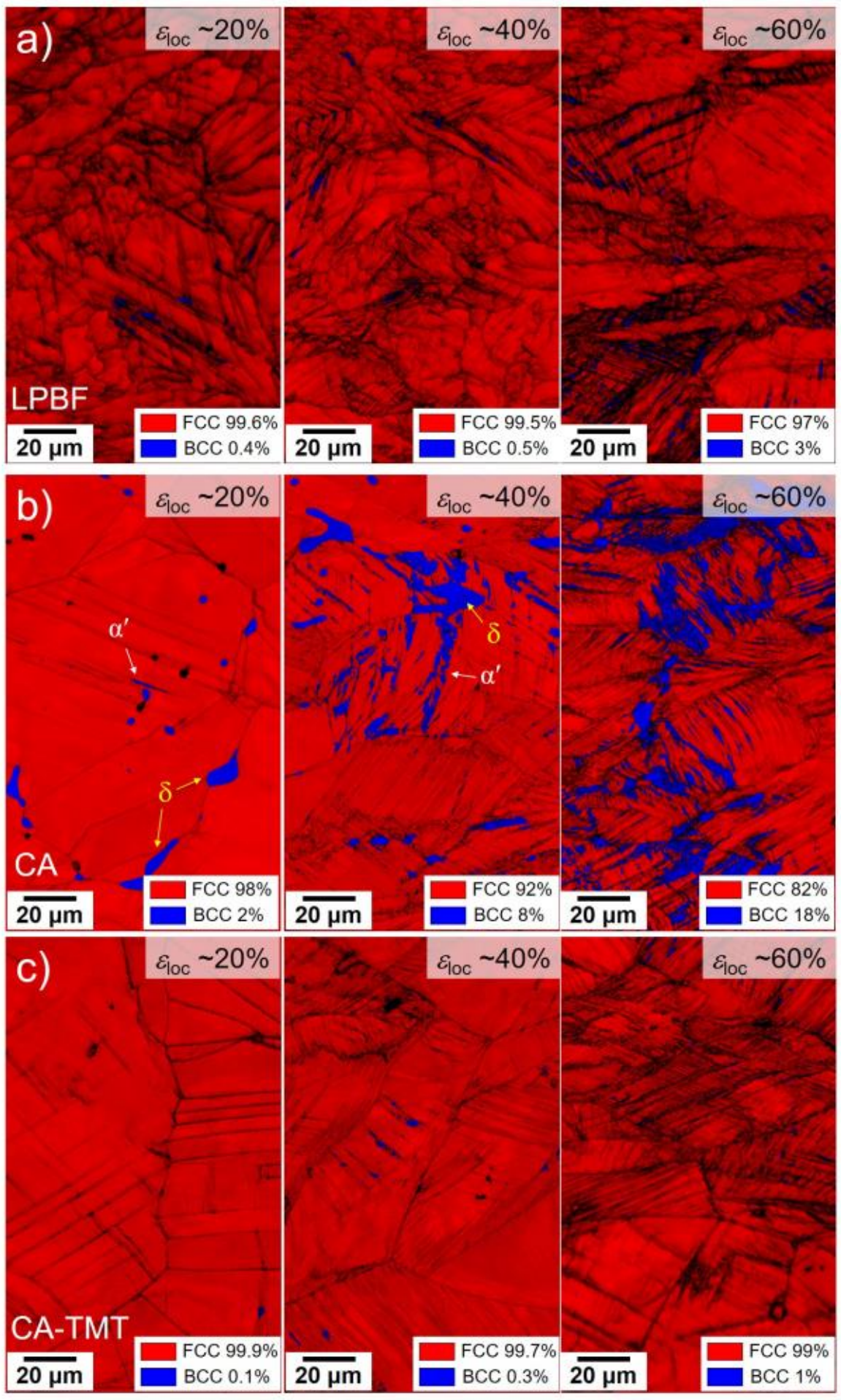

Fig. 7 EBSD phase map with IQ map superimposed of the (a) LPBF, (b) CA, and (c) CA-TMT samples with increasing local strain from ~20% to ~60%. (LPBF: Laser powder-bed-fusion, CA: Casting plus annealing, TMT: Thermomechanical treatment).

## 3.3 H-induced fracture and cracking

Fig. 8 shows high- and low-magnification fractographs of the tensile tested specimens of the 5 days H-charged samples. Since H effects on ductility are more clearly apparent in the 5 days charged samples (especially for samples LPBF and CA-TMT), the investigation of H-induced fracture and cracking was focused on the 5 days charged samples. Thus, the images of the H-charged samples shown hereafter are those of the 5 days charged material. The fracture surfaces of the H-charged specimens consist of two regions with distinctly different features. The central regions of the fracture surface show dimples (Fig. 8$a_2$,$b_2$,$c_2$), whereas the regions close to the side edges of the sample, where the material was exposed to H-charging, exhibit the characteristics of quasi-cleavage fracture (Fig. 8a3,b3,c3). Generally, when electrochemical H charging is conducted, especially on FCC metals and samples which have a relatively low H diffusivity, H is highly concentrated near the charged surface and thus a H concentration gradient from the edge to the inner regions of the samples is induced [5,8]. For this reason, the quasi-cleavage fracture type observed at the specimen edges is attributed to the high H concentration. An interesting feature can be found when comparing the H-induced brittle regions among the samples (Fig. 8a3,b3,c3): the depth of the brittle region, i.e. the H-affected zone, in samples CA-TMT and LPBF is significantly shallower (~20–50 μm) than that in sample CA (~140–180 μm). This observation further supports that samples CA-TMT and LPBF are less susceptible to HE than the CA sample.

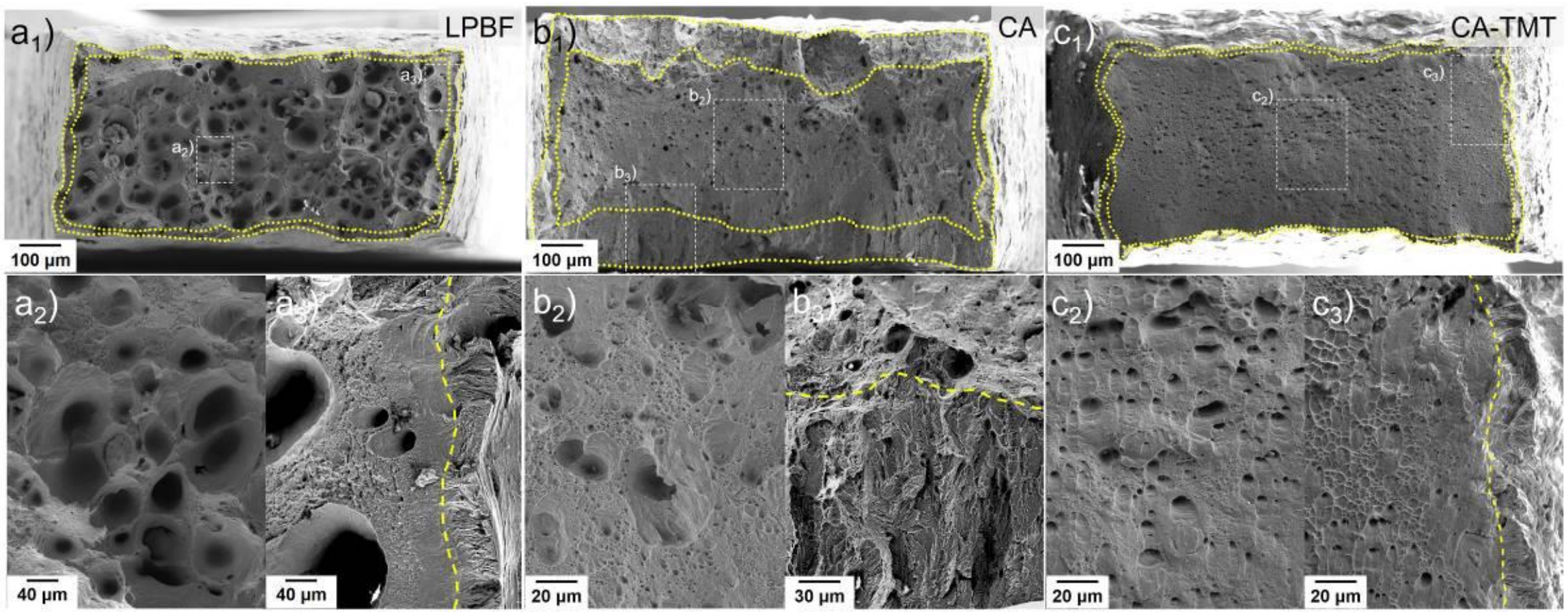

Fig. 8 The low- and high-magnification SEM images of the tensile fracture surfaces for H-charged the (a) LPBF, (b) CA, and (c) CA-TMT samples. (LPBF: Laser powder-bed-fusion, CA: Casting plus annealing, TMT: Thermomechanical treatment).

To understand the H-assisted cracking behavior, cross-sections of the tensile-tested specimens were cut along the tensile direction and observed by optical microscopy first. Fig. 9 displays representative optical micrographs of the cross-sectional region in the tensile-tested sample near the fracture surface for all samples.

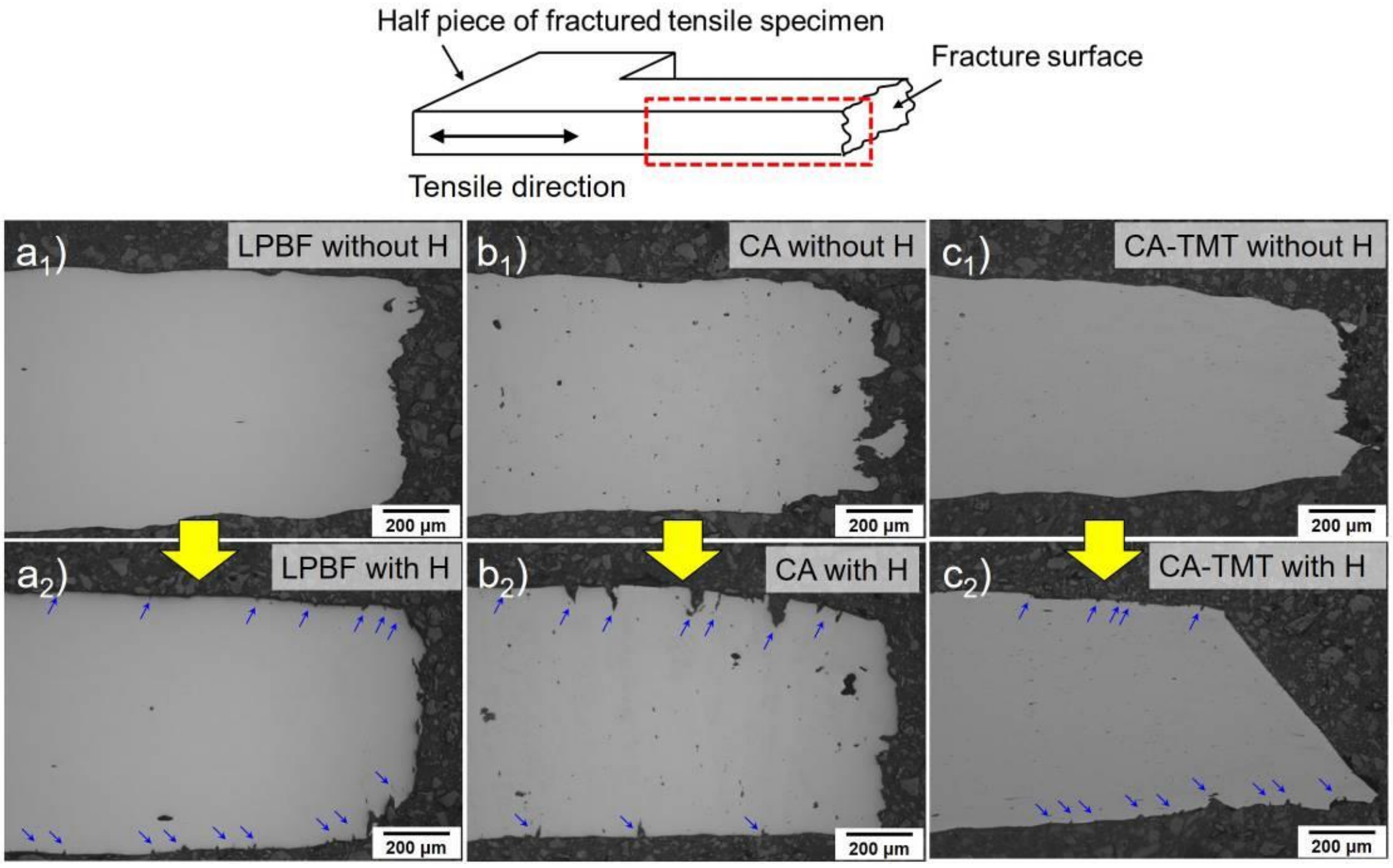

Fig. 9 Optical micrographs of the cross-sectional region in the tensile-tested sample near the fracture surface. The arrows indicate the H-assisted cracks. (LPBF: Laser powder-bed-fusion, CA: Casting plus annealing, TMT: Thermomechanical treatment).

Two important features are noteworthy in Fig. 9. First, only the H-charged samples show surface cracks propagating perpendicular to the tensile axis. Second, it is clear that the depth of these H-assisted cracks is much shallower in samples LPBF and CA-TMT compared with the CA sample, proving a higher resistance to H-assisted cracking in the former. To identify the correlations between the microstructure and the crack propagation path, more detailed microstructural observations were conducted on the secondary cracks by means of EBSD measurements. Fig. 10a–c show the image quality (IQ) map and the corresponding phase map for the H-charged CA, CA-TMT, and LPBF samples, respectively. It is revealed that all samples show similar H-assisted cracking behavior; the majority of cracks observed nucleate and propagate along the γ-α′ interface (denoted γ-α′ interface crack) as well as within α′ (denoted α′ crack) as indicated by white and blue arrows in Fig. 10, respectively. Rarely, H-assisted cracks are propagating through both γ and α′. A high magnification image of these cracks (Fig. 10$a_3$,$b_3$,$c_3$) suggests that these alternating cracks are formed by coalescence of several γ-α′ interface cracks and α′ cracks. In some cases, the cracks tend to initiate at the γ-α′ interfaces (or within α′) and propagate through the γ matrix, as indicated by yellow arrows in Fig. 10. However, since the crack propagation through austenite only occurs in narrow γ layers between α′ martensite regions as a coalescence process of α′-related cracks, it presumably has only a minor effect on the overall H-assisted cracking behavior. Thus, it is apparent that the strain-induced martensitic transformation plays an important role in H-assisted cracking for all types of samples.

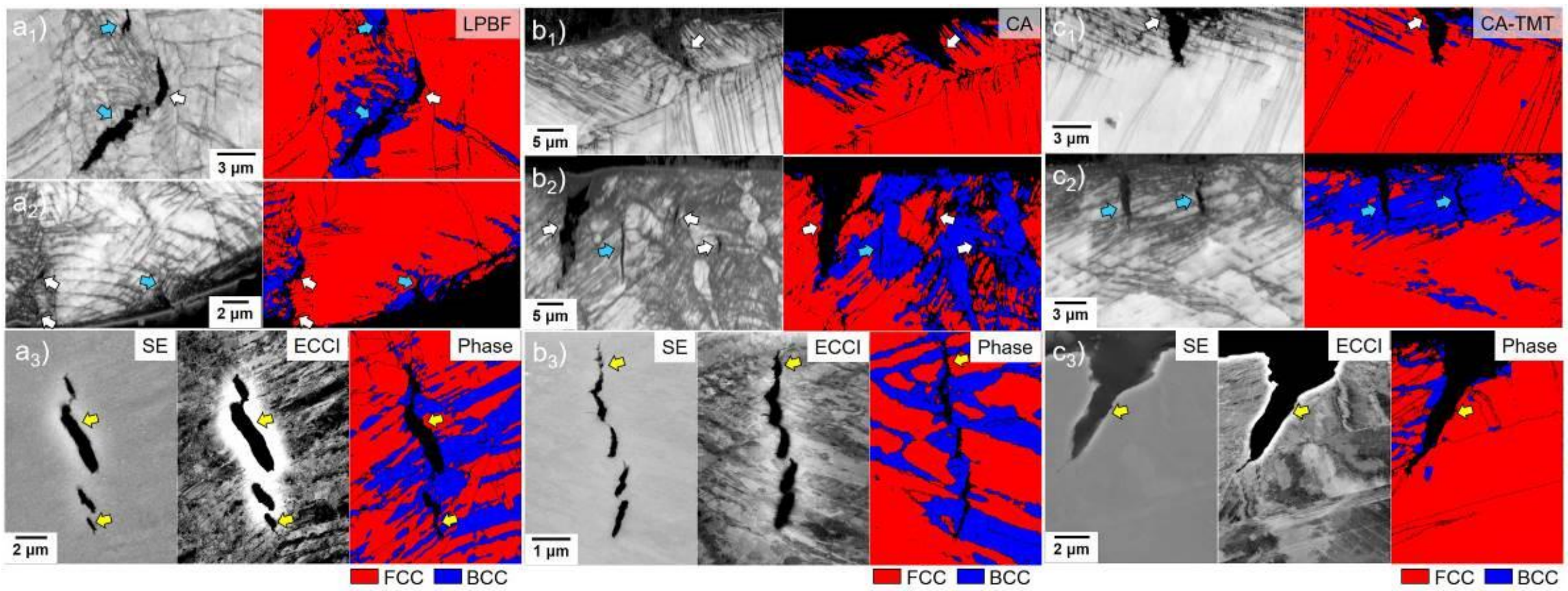

Fig. 10 Secondary H-assisted cracks in samples (a) LPBF, (b) CA, and (c) CA-TMT: ($a_{1,2}$)–($c_{1,2}$) show IQ maps and the corresponding phase distribution maps obtained from EBSD measurement around the H-assisted cracks; ($a_3$)–($c_3$) show high magnification secondary electron (SE) images and the corresponding ECCI and EBSD phase map of the H-assisted cracks. The white, blue and yellow arrows denote γ-α′ interface crack, α′ crack, and γ penetrating cracks, respectively. (LPBF: Laser powder-bed-fusion, CA: Casting plus annealing, TMT: Thermomechanical treatment).

## 4. Discussion

Results of the tensile experiments in Section 3 clearly demonstrate that HE in 304L ASS is apparently related not only to the presence of retained δ ferrite, but also to the amount of strain-induced α′ martensite. However, the role of δ ferrite in HE of ASSs still remains controversial, although it has been studied by many researchers [14,15,47,48]. Moreover, some studies reported that there was no significant effect of δ on HE when the amount of δ is less than 10% [15,47]. In this study, only ~1% δ is observed in sample CA and, regardless of the presence of δ, all investigated samples show similar H-assisted cracking behavior, i.e. cracks are initiated in the region where strain-induced martensitic transformation occurs. Even in the CA sample, H-assisted cracks are more likely formed near strain-induced α′ rather than at the retained δ (see Fig. $10b_2$). Thus, it can be suggested that strain-induced martensitic transformation is the primary factor controlling the HE susceptibility in 304L ASSs and the retained δ may partially contribute to this process.

Therefore, in the case of 304L ASS, it is important to achieve high austenite stability for likewise achieving high HE resistance. For this purpose, 304L produced by conventional casting requires several thermomechanical treatments, whereas LPBF-processed 304L does not require any further processing steps. From this point of view, it can be concluded that the LPBF process itself leads to enhanced strength in the ASS, and it also renders the sample less or equally susceptible to HE compared to conventionally produced 304L. Below, we discuss these results in terms of variations in mechanical stability of austenite, H-assisted cracking behavior, and H diffusion and trapping behavior.

### 4.1 Variations in mechanical stability of austenite

First of all, it is important to understand why the CA sample shows a much lower austenite stability than the others, because this observation deviates from what is expected from the $\Delta G^{\gamma\rightarrow\alpha}$ and $M_{d30}$ values. Considering that $\Delta G^{\gamma\rightarrow\alpha}$ and $M_{d30}$ values are calculated based on overall chemical composition, this inconsistency may arise from a possible local chemical inhomogeneity in the samples. To check this possibility, the elemental distribution was analyzed by EDS elemental mapping (Fig. 11 and Fig. 12). The back-scattered electron (BSE) image and the corresponding EDS maps of sample LPBF (Fig. 11a) indicate a homogeneous distribution of constituent elements on the micro-scale. STEM annular dark field (ADF) image shows a high density of the dislocation structure near the cell wall, however, we could not find any indication of elemental segregation or clustering from the STEM-EDS elemental mapping (Fig. 11b). Thus, we can conclude that the constituting elements are homogeneously distributed in sample LPBF despite the presence of a solidification cellular structure. This observation is in good agreement with the results obtained in rapid solidification experiments including splat quenching and laser welding which showed that a higher cooling rate leads to partitionless austenitic solidification [49]. Thus, the high characteristic cooling rate associated with the LPBF process results in more homogeneous elemental distribution compared to the conventional processing routes encountered in commercial production of this steel. However, the EDS results of the CA sample (Fig. 12a) reveals that elemental partitioning between γ and δ is evident for Cr and Ni and, more importantly, the γ regions nearby δ are enriched in Cr but depleted in Ni. This clear chemical inhomogeneity observed in the γ matrix may be attributed to the incomplete partitioning of Cr and Ni to the γ that had transformed from the δ during casting. The detailed mechanism on this phenomenon and its verification are provided in the Supplementary Material. Since Cr and Ni are considered as ferrite stabilizer and austenite stabilizer, respectively, the γ region nearby δ would exhibit a low austenite stability. In contrast to the CA sample, a homogeneous elemental distribution is observed in the CA-TMT sample (see Fig. 12b), suggesting that the imposed TMT completely eliminates chemical inhomogeneity as well as δ from the CA sample, and thus enhances its austenite stability.

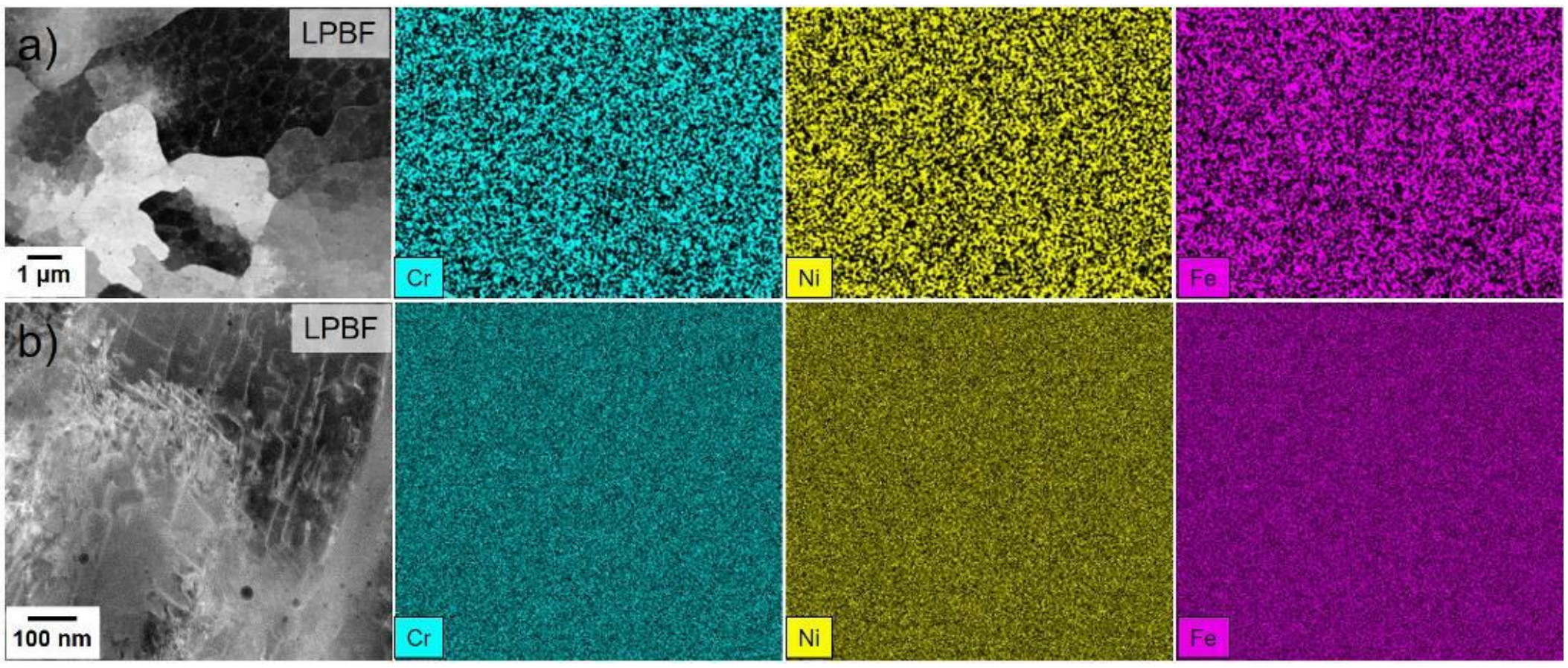

Fig. 11 Elemental distribution in sample LPBF. (a) BSE image and the corresponding EDS maps and (b) STEM-ADF image and the corresponding EDS maps. (LPBF: Laser powder-bed-fusion).

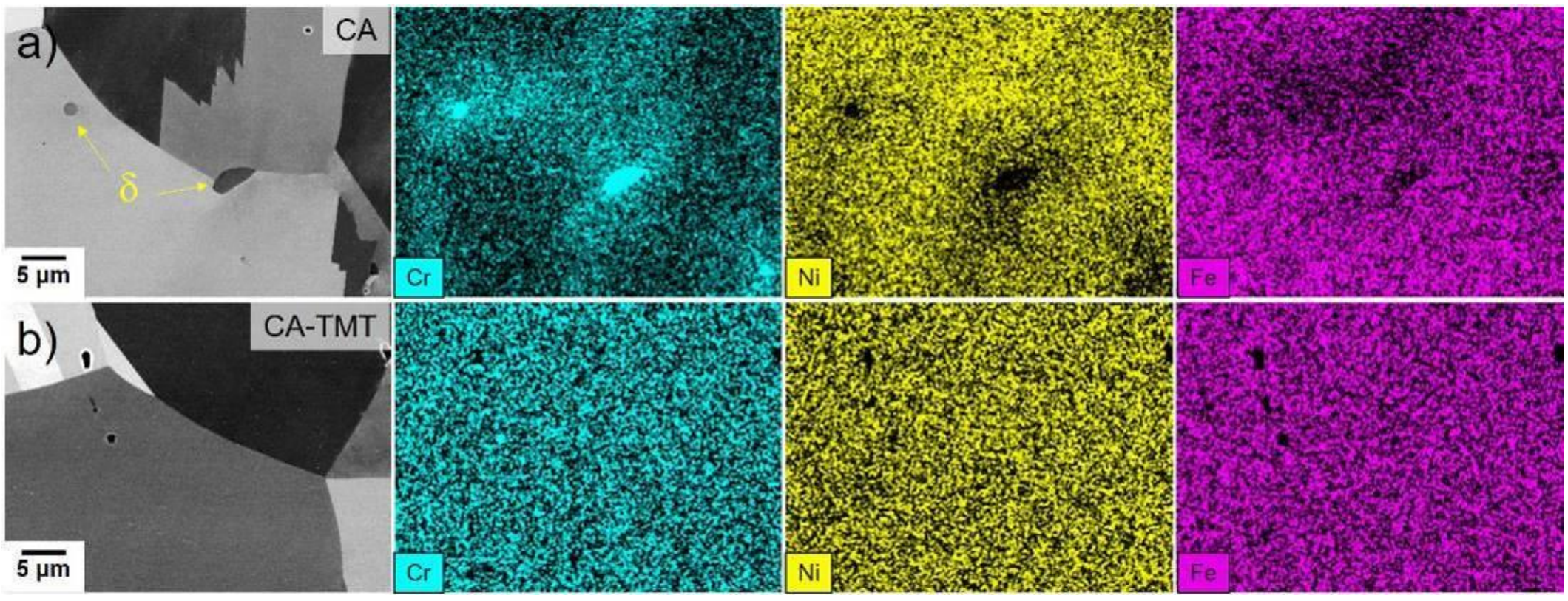

Fig. 12 BSE image and the corresponding EDS maps of sample (a) CA and (b) CA-TMT. (CA: Casting plus annealing, TMT: Thermomechanical treatment).

In addition to the chemical matrix inhomogeneity, it is often found in the deformed CA sample that strain (and the associated dislocation density) is localized at γ-δ interfaces, as shown in Fig. 13. In general, a steep strain gradient is present near the interfaces between the soft austenite matrix and hard inclusion phases, translating into a higher density of geometrically necessary dislocations (GNDs) in the soft matrix [50–53]. Since dislocations are well known to lower the barrier for strain-induced martensitic transformation [54], the increased GND density near the γ-δ interface conceivably contributes to α′ formation. Thus, a large volume fraction of α′ in sample CA is mainly attributed to the chemical inhomogeneity-induced localized reduction of austenite stability, and the presence of δ additionally contributes to the strain-induced martensitic transformation by inducing a high density of GNDs.

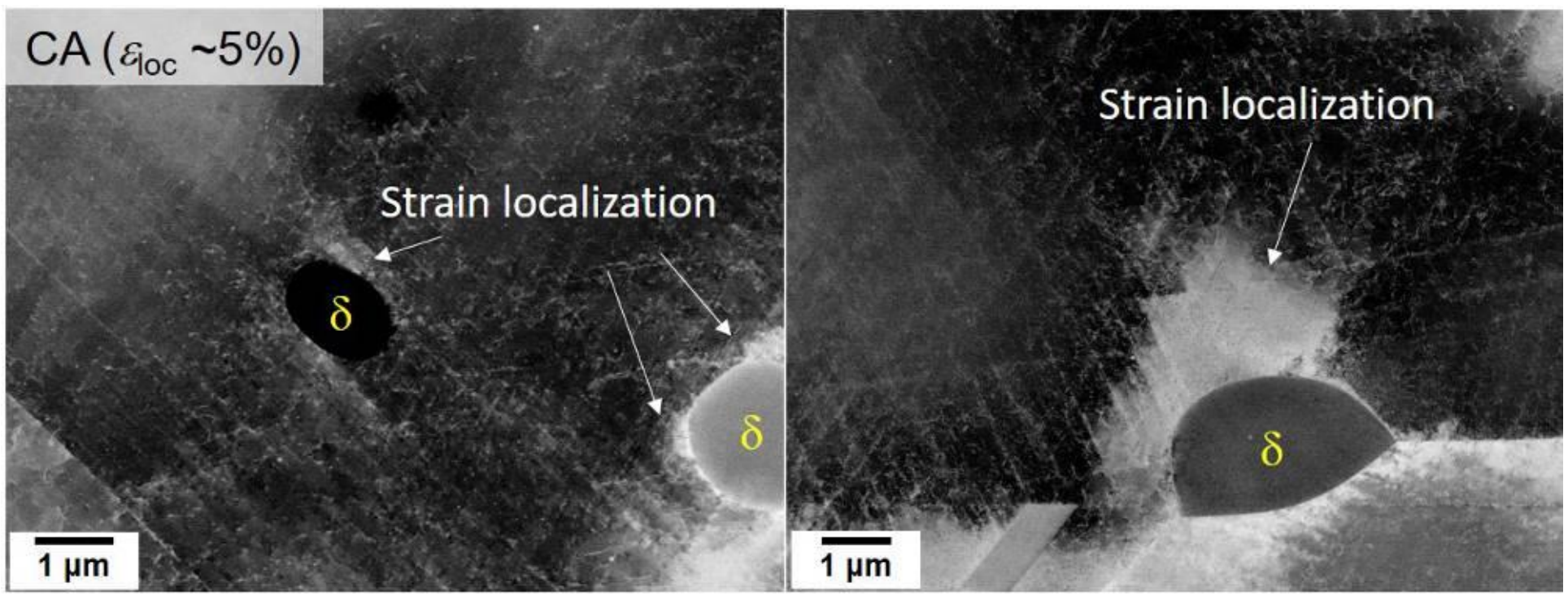

Fig. 13 ECCI images of deformed the CA sample showing strain localization near the retained δ. (CA: Casting plus annealing).

Another interesting feature to be noted is that although the LPBF sample clearly exhibits a cellular structure in its initial state (after solidification), its austenite stability is similar to that of the CA-TMT sample. In fact, $f_{\alpha'}$ in sample LPBF is slightly higher than that in the CA-TMT sample for all investigated strain levels (see Fig. 7). However, this is simply due to the fact that the driving force ($\Delta G^{\gamma\rightarrow\alpha}$) is slightly higher in sample LPBF than in sample CA-TMT. Thus, it can be suggested that neither nucleation nor growth of α′ is significantly affected by the presence of the solidification cell substructure. With regard to the growth of α′, one may argue that the solidification cell substructure in the LPBF sample may act as a barrier that hinders the growth of α′, since this structure is stable during the entire deformation. However, this possibility can be readily ruled out because it is often found that the size of α′ martensite is bigger than that of the cellular substructure. The combined EBSD and ECCI observations of regions containing α′ martensite (Fig. 14) suggest that strain-induced α′ can easily pass through adjacent cells even at relatively low strain level (e.g., ~20%). Since the dislocation cell wall is analogous to LAGBs rather than to HAGBs, it might have a minor effect on suppressing the growth of α′ [23].

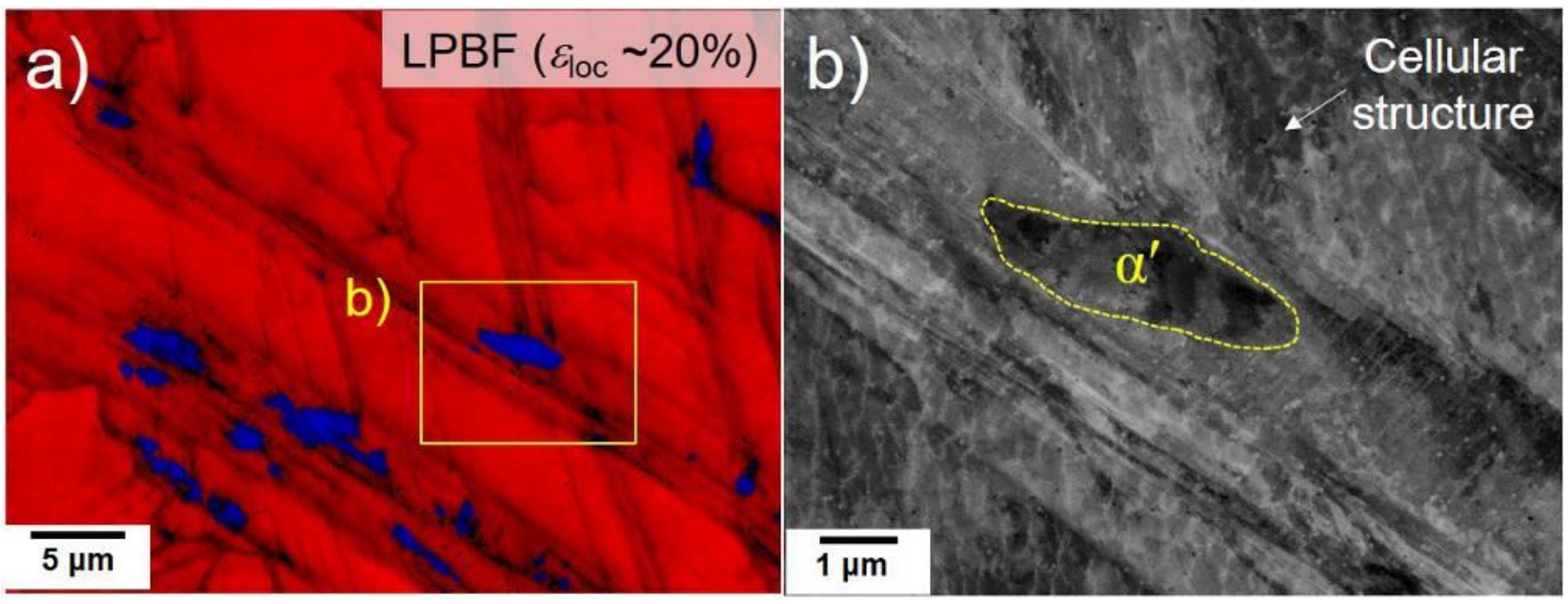

Fig. 14 Distribution of strain-induced α′ martensite in sample LPBF strained to ~20%: (a) phase distribution map and (b) magnified ECCI image of a selected area highlighted in (a). (LPBF: Laser powder-bed-fusion).

## 4.2. H-assisted cracking mechanism

### 4.2.1 Initiation of H-assisted cracking

Before identifying the mechanism of H-assisted crack initiation, the sequence of the H-assisted cracking processes should be discussed first, since one could argue that H-assisted cracks are first initiated within γ, and then the high strain localization near the crack tip promotes the martensitic transformation. However, considering that stresses are relaxed upon crack initiation, the microstructure in the vicinity of the crack initiation site (perpendicular to the crack propagation path) should represent the condition at the time of crack nucleation. In this study, the majority of crack initiation sites, which is typically located in the center of the crack, are identified as the γ-α′ interfaces and the inside regions of α′. Thus, it is likely that strain-induced martensitic transformation occurs prior to H-assisted cracking instead of being a result of cracking. More importantly, there is no evidence that H-assisted cracks and/or voids are initiated at the critical sites of γ, such as GBs and intersections of slip lines (or mechanical twins), as shown in ECCI images of Fig. 10. This supports our hypothesis that, at least for the present samples, strain-induced martensitic transformation occurs first, and then the regions containing α′ act as the crack initiation sites.

In fact, strain-induced martensitic transformation is a crucial factor in the severe HE of metastable ASS despite some dispute in the literature [9,42,55]. The general consensus is that the strain-induced martensite contributes strongly to H-assisted cracking [11,45,56], which can be explained as follows: in general, the body centered cubic (BCC) structure of α′ martensite is inherently more susceptible to H-assisted cracking than the FCC austenitic structure [9,10,57]. Furthermore, the H solubility in α′ martensite is an order of magnitude lower than that in austenite [58]. Thus, when H-containing austenite is transformed to α′ martensite and the martensite inherits the H from its host austenite matrix, it suddenly exceeds the solubility limit in α′ martensite, i.e. the H becomes excess H. Following the transformation, excess H diffuses to the surrounding austenite and accumulates at the γ-α′ interphase boundary, since the H diffusivity is much higher in α′ martensite than in austenite [58,59]. When a critical H concentration is attained at this boundary, H leads to decohesion at the γ-α′ interphase boundary and thus cracks start to nucleate and propagate along this boundary (γ-α′ interface crack) [12,60,61]. Furthermore, H-enhanced slip localization in the retained γ can cause void formation [62,63] along the γ-α′ interface, further facilitating the decohesion process. Additionally, the excess H can also be accumulated within the substructure of α′ martensite (i.e., at lath/block boundaries). In this case, supersaturated H at lath/block boundaries might lead to decohesion and to crack initiation/propagation within α′ martensite (α′ crack) [45,64,65]. Since both, γ-α′ interface cracks and α′ cracks are frequently observed in Fig. 10, it seems that all mechanisms mentioned above are activated. Thus, the excess H resulting from the strain-induced α′ martensitic transformation plays a crucial role in HE of the present samples. In this regard, it can be concluded that the presence of the austenite with low stability around δ in the CA sample enhances the localized strain-induced α′ martensite transformation, which in turn leads to the formation of possible crack initiation sites in the presence of H, and thus contribute to the severe HE susceptibility.

#### 4.2.2. Growth of H-assisted cracking

In addition to the crack initiation, the crack growth resistance is another crucial factor to control the elongation to failure in a H-providing environment; i.e., low resistance to crack growth results in brittle fracture at low strain levels. In this study, it was clearly observed that the CA sample shows a much higher number of deep H-assisted cracks than the other samples (Fig. 9), indicating that the CA sample exhibits a lower resistance to crack growth when exposed to a H-donating atmosphere.

It has been suggested that the H diffusion ahead of a crack tip plays a key role in the growth of H-assisted cracks. In tensile loading condition, a stress field with a hydrostatic component is developed in front of a crack tip and thus the enrichment of H at a critical site ahead of the crack tip can be attained by stress-driven H diffusion [61,66–68]. When the accumulated H concentration exceeds the critical value near the crack tip, H leads to continuous crack growth (new cracks initiate and coalesce with the pre-existing crack) and finally causes the complete fracture of the material [69,70]. Thus, it can be concluded that the acceleration in crack growth due to H depends on the rate at which the H is provided to the critical region ahead of crack tip and thus to the local H concentration. On the other hand, it might be reasonable to assume that H diffusion to a crack tip can be accelerated by the presence of α′ near the crack, due to the high diffusion rate of H in α′ [58,59]. Besides the pre-existing α′, fresh α′ can be formed in the stress field in front of the crack tip, leading to further enhancement of H diffusion [18,61] and providing a high local oversaturation of H. In fact, Wang et al. [61] investigated the influence of martensitic transformation in front of a crack tip on the H enrichment by means of finite element analysis. They reported that the enrichment of H in front of the crack tip is accelerated by about four orders of magnitude due to the presence of α′ at the crack tip. Since the amount of both pre-existing α′ and fresh α′ around the crack is dependent on the austenite stability, a correlation between the austenite stability and the capacity of H diffusion ahead of the crack can be derived. In addition to H diffusivity, austenite stability may affect the stress field at the crack tip. That is, the α′ formation in front of a crack tip may reduce the hydrostatic tensile stress, $\sigma_h$, at the crack tip, since a compressive stress, $\sigma_c$, can be created in the transformed region due to the volume change associated with the phase transformation [71]. The reduction in $\sigma_h$ at the crack tip not only enhances the crack growth resistance but also decreases the driving force for H diffusion toward the crack tip. However, with decreasing austenite stability, the reduction in $\sigma_h$ at the crack tip is gradually diminished due to the smaller amount of retained γ surrounding the crack tip (i.e., smaller transformed volume and thus $\sigma_c$) [71]. In this respect, it is reasonable to assume that the reduction in driving forces for both crack growth and H diffusion to the crack tip can be more pronounced in more stable austenite, compared to less stable austenite. Based on these concepts, a relationship between austenite stability and resistance to crack propagation can be derived, as schematically illustrated in Fig. 15. In the CA sample, which has locally a low austenite stability, a large amount of α′ near the H-assisted crack significantly accelerates the stress-driven H diffusion to the crack tip and thus leads to continuous crack growth (see Fig. 15a). On the other hand, in the LPBF and CA-TMT materials, which have high austenite stability, the enrichment of H at the crack tip is not sufficient due to the small amount of α′, and thus the H-assisted cracks can be easily arrested (see Fig. 15b). We therefore conclude that in the CA sample, a high capacity of H diffusion toward the crack tip leads to the continuous growth of H-assisted cracks and thus further contributes to material's HE susceptibility. For the same reason, the depth of the H-affected brittle zone in the CA samples is significantly deeper than that in the LPBF and CA-TMT specimens (Fig. 8).

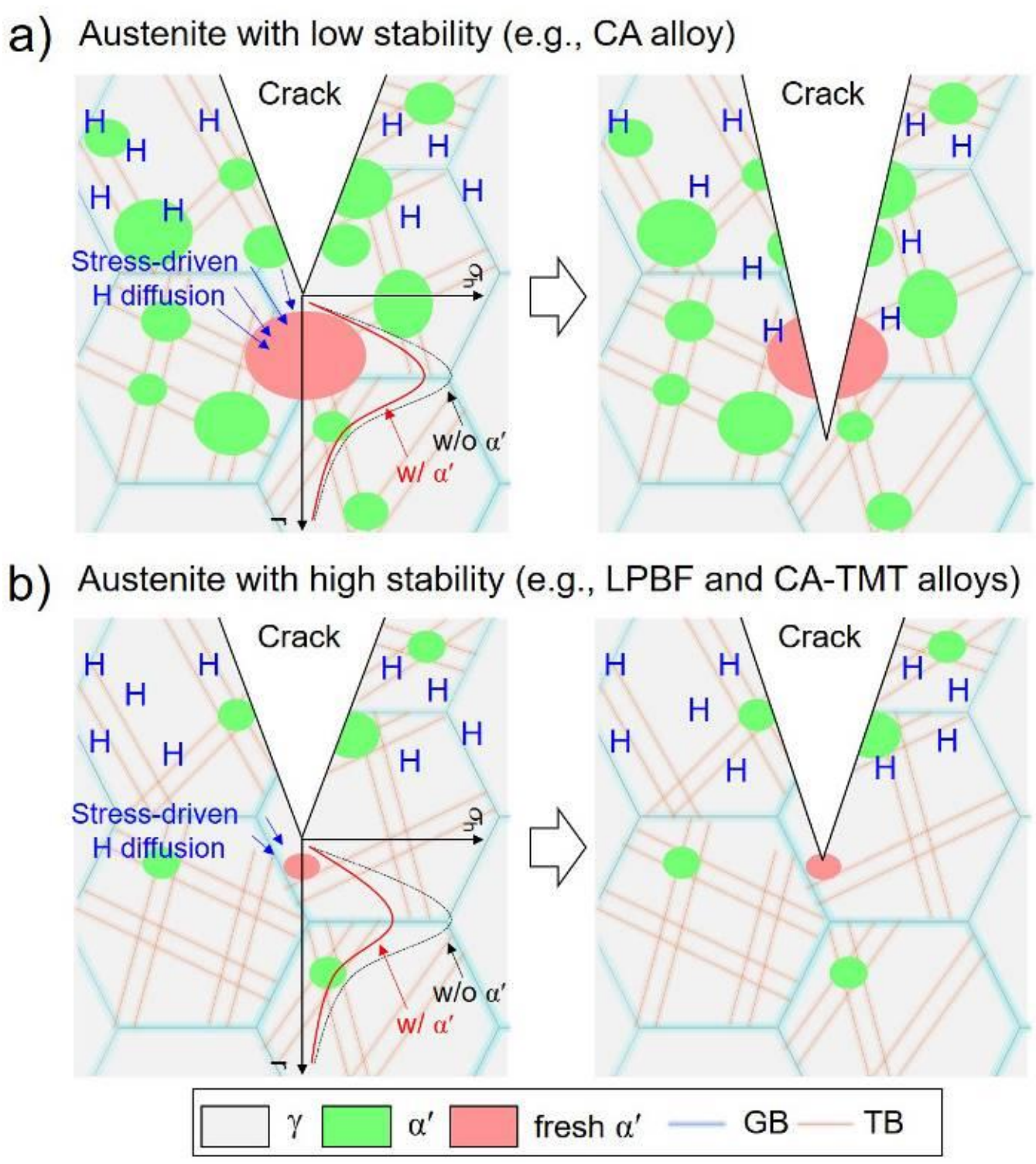

Fig. 15 Schematic illustration showing the influence of stress-driven H diffusion on crack growth resistance for austenite with (a) low stability and (b) high stability. Note that the shape of α′ is simplified as ellipsoid, but its actual shape is rodlike. (GB: Grain boundary, TB: Twin boundary).

### 4.3 Variations in H diffusion and trapping behavior

Since one may argue that the depth of a H-assisted crack (or of the H-affected brittle zone) is closely related to the H permeation depth after H-charging, it is instructive to investigate the possible difference in H permeation depth after H-charging among the investigated samples. Under the assumption that the H lattice diffusion is largely controlled by Fick's law, the local H content, $C_H$, as a function of the depth from the surface, $x$, can be estimated as [72]:

$$C_H(x,t) = C_{Hs}\left[1 - erf\left(\frac{x}{2\sqrt{D_H t}}\right)\right], \tag{2}$$

where $C_{Hs}$ is the local H content at the surface of the H-charged samples, $t$ is the charging time, and $D_H$ is the diffusion coefficient of H. In Eq. (2), it is evident that when $x/(2\sqrt{D_H t}) \geq 2$, $C_H$ approaches 0, hence, the maximum H permeation depth, $x_{max}$, after H-charging can be estimated by taking $x_{max}/(2\sqrt{D_H t}) = 2$, i.e., $x_{max} = 4\sqrt{D_H t}$. Since all samples were H-charged under the same conditions, $x_{max}$ only depends on $D_H$. One could expect that the cell structure with high-density of dislocation in the LPBF sample reduces the H diffusivity, since dislocations act as H trapping sites. However, in contrast to BCC metals, the increase in dislocation density has little influence on the apparent H diffusivity in ASSs [16,20,73], because the binding energy of H at dislocations in FCC metals is lower than the activation energy for H lattice diffusion, whereas in BCC metals the binding energy between H and the lattice dislocations is much higher than the activation energy for H lattice diffusion [20]. In this regard, $D_H$ for all investigated samples can be assumed to be the same as that of 300-series ASS (~$3.17\times10^{-16}$ $m^2/s$ at RT [3]). By taking this $D_H$ value, the $x_{max}$ after 5 days H-charging can be estimated as ~47 μm for all investigated samples. As shown in Fig. 8 and Fig. 9, however, a much higher depth of the H-affected brittle zone (or of the H-assisted crack) was measured from the analysis of fracture surfaces in the CA sample (~180 μm). The discrepancy between $x_{max}$ and the depth of the H-affected brittle zone suggests that further permeation of H atoms occurs during the deformation. The reason for this may be stress-driven H diffusion toward the crack tip, as discussed above. From this viewpoint, it can be accepted that the depth of the H-affected brittle zone in samples LPBF and CA-TMT is similar to the estimated $x_{max}$, since both samples show a low capacity of H diffusion toward the crack tip. Note that the CA sample contains retained δ and adjacent austenite regions enriched in Cr and depleted in Ni, but the amount of retained δ is too small to affect the H diffusion and the high Cr content is known to have a rather negative effect on the H diffusion [3]. Therefore, this observation further supports our hypothesis that stress-driven H diffusion toward the crack tip plays a crucial role in H-assisted crack growth. Although the motion of dislocations with trapped H can carry H deep into the material and thus assist the H atoms' diffusion process [74,75], the shallow depth of the H-affected brittle zone in the LPBF and CA-TMT materials, which is similar to the estimated depth of H permeation, implies that mobile dislocations have a negligible effect on H diffusion in the present samples.

In contrast to the H diffusivity, the H content in ASSs increases with the density of dislocations, since dislocations provide additional H trapping sites [3,20]. However, TDS results in Fig. 5 reveal that the solidification cell substructure, which is decorated with dislocations, does not increase the H content of the LPBF sample significantly. Thus, it seems that, in contrast to the dislocations, the solidification cell structure may not provide additional H trapping sites and thus increase the H content. Although the detailed mechanism is not fully understood yet, it is reasonable to suggest that the low H trapping ability of the solidification cell structure is due to its low binding energy to H atoms. Since the solidification cell structure in the LPBF sample possesses a low-energy dislocation structure [31,76], the elastic stress field around this structure is different from that around the randomly distributed dislocations [77]. As a result, the binding energy of H to the solidification cell also might be different from that of randomly distributed dislocations. However, it is not feasible to investigate the influence of the solidification cell structure on the activation energy for H desorption under the present experimental conditions. For example, the TDS results in Fig. 5 show only one low-temperature peak, but there is a possibility that several peaks, which correspond to various weak H trapping sites such as GBs and dislocations, may be overlapped. Therefore, additional systematic experiments and simulations are now required to confirm the present trend and to provide information on the precise mechanism of the low H trapping ability of the solidification cell structure.

## 5. Conclusions

In the present study, H effects on tensile ductility of 304L ASS fabricated by laser powder-bed-fusion (LPBF) were systematically investigated through a series of tensile tests, microstructural analyses and by comparing the results with those of conventionally produced 304L ASSs (i.e., cast and annealed (CA) and cast, annealed and thermomechanically-treated (CA-TMT) samples). The major conclusions of this investigation are as follows:

1. Hydrogen embrittlement susceptibility of 304L ASSs varied with different production routes; H-charging resulted in a significant loss in ductility for the CA sample, whereas only a small effect of H on ductility was observed for samples LPBF and CA-TMT. In the EBSD analysis, the volume fraction of strain-induced α′ martensite was higher in the CA sample (~16%) than in samples LPBF and CA-TMT (~3% and ~1%, respectively) at the same strain level of 60%, suggesting that the low local mechanical stability of austenite may be the reason for the low resistance to hydrogen embrittlement in the CA sample.
2. The low mechanical stability of austenite in sample CA can be explained by the presence of chemical inhomogeneity in austenite regions particularly around the retained δ zones in the initial state. Thermomechanical treatment, subsequent to CA, enhanced the chemical homogeneity, which in turn can effectively recover the austenite's stability and thus the resistance to hydrogen embrittlement in the CA-TMT sample.

3. In contrast, the LPBF sample showed a fully austenitic structure with homogeneous distributions of its constituent elements and a highly complex microstructure containing a characteristic solidification cellular structure due to the high solidification and cooling rates. These features led not only to enhanced strength but also to a high resistance to hydrogen embrittlement in the ASS without any further processing step.
4. In all samples, most of the H-assisted cracks were initiated at regions with α′ martensite, which further supports the critical role of strain-induced α′ martensitic transformation in hydrogen embrittlement of the present samples. Thus, it was revealed that the presence of δ and the chemical inhomogeneity inside γ matrix promoted hydrogen embrittlement by enhancing the deformation-induced martensitic transformation and the associated H enrichment at the γ-α′ interface.
5. The LPBF sample, in spite of the presence of a solidification cellular structure, showed similar HE susceptibility to the CA-TMT sample. This can be rationalized by the negligible effect of the solidification cell structure on austenite stability as well as on H trapping behavior.

**Acknowledgements**

The authors would like to gratefully acknowledge P. Beley for the help of TDS analysis and the kind support of M. Nellessen, K. Angenendt, and M. Adamek at the Max-Planck-Institut für Eisenforschung. D.-H. Lee gratefully acknowledges the research fellowship provided by the Alexander von Humboldt Stiftung (AvH, Alexander von Humboldt Foundation, https://www.humboldt-foundation.de).